\documentclass[aps,pra,reprint,superscriptaddress,showkeys]{revtex4-1}
\usepackage{graphicx}
\usepackage{amsmath}
\usepackage{textcomp}
\usepackage{color}
\usepackage{xcolor}
\usepackage{upgreek}
\usepackage{comment}
\usepackage{verbatim}
\usepackage{amssymb}
\usepackage{tabu}
\usepackage{float}
\usepackage[english]{babel}

\begin{document}

\title{Mirror-Symmetric Heterogeneous Resonant Nanostructures: Extrinsic Chirality and Spin-Polarized Scattering}

\author{Sergey Nechayev}
\affiliation{Max Planck Institute for the Science of Light, D-91058 Erlangen, Germany}
\affiliation{Institute of Optics, Information and Photonics, Friedrich-Alexander-University Erlangen-Nuremberg, D-91058 Erlangen, Germany}

\author{Pawe{\l} Wo{\'z}niak}
\affiliation{Max Planck Institute for the Science of Light, D-91058 Erlangen, Germany}
\affiliation{Institute of Optics, Information and Photonics, Friedrich-Alexander-University Erlangen-Nuremberg, D-91058 Erlangen, Germany}

\author{Martin Neugebauer}
\affiliation{Max Planck Institute for the Science of Light, D-91058 Erlangen, Germany}
\affiliation{Institute of Optics, Information and Photonics, Friedrich-Alexander-University Erlangen-Nuremberg, D-91058 Erlangen, Germany}

\author{Ren{\'e} Barczyk}
\affiliation{Max Planck Institute for the Science of Light, D-91058 Erlangen, Germany}
\affiliation{Institute of Optics, Information and Photonics, Friedrich-Alexander-University Erlangen-Nuremberg, D-91058 Erlangen, Germany}


\author{Peter Banzer}
\email[]{peter.banzer@mpl.mpg.de}
\homepage[]{http://www.mpl.mpg.de/}
\affiliation{Max Planck Institute for the Science of Light, D-91058 Erlangen, Germany}
\affiliation{Institute of Optics, Information and Photonics, Friedrich-Alexander-University Erlangen-Nuremberg, D-91058 Erlangen, Germany}
\date{\today}

\begin{abstract}
{We investigate a geometrically symmetric gold-silicon sphere heterodimer and reveal its extrinsic chiroptical response  caused by the interaction with a substrate. The chiroptical response is obtained for oblique incidence owing to the coalescence of extrinsic chirality, heterogeneity and substrate induced break of symmetry. To quantify the chiral response we utilize k-space polarimetry. We elucidate the physics of the involved phenomena by considering scattering properties of the heterodimer in free space and find that incident linearly polarized light is scattered in a spin-split fashion. We corroborate our finding with a coupled dipole model and find that the spin-split behavior originates from the heterogeneity of the structure. This spin-split scattering, combined with the substrate-induced break of symmetry, leads to an extrinsic chiroptical response. Our work sheds new light on the potential and optical properties of heterogeneous nanostructures and paves the way for designing spectrally tunable polarization controlled heterogeneous optical elements.}
\end{abstract}
\keywords{chirality, bianisotropy, heterodimer, scattering, coupled-dipole, magneto-electric, meta-atoms}
\maketitle
%
\section*{Introduction} 
\vspace{-0.5cm}
Chirality refers to the geometry of objects lacking in mirror-symmetry~\cite{kelvin_baltimore_1904,barron_parity_1972}. A chiral object can exist in two forms of opposite handedness, called enantiomers. Enantiomers can be distinguished optically, for instance, via their interaction with circularly polarized light (CP)~\cite{tang_optical_2010}. Artificial chiral nanostructures are typically made of metallic wires following a helical path in space~\cite{hoflich_direct_2011}. These structures exhibit a stronger chiroptical response than that of chiral molecules, owing to larger induced electric and magnetic dipole moments~\cite{fan_plasmonic_2010,schaferling_tailoring_2012}. Chiral structures can also be realized by arranging achiral building blocks such as disks and spheres to form chiral compositions~\cite{fan_plasmonic_2010,hentschel_three-dimensional_2012,hentschel_optical_2012,hentschel_plasmonic_2013}.
\\
An alternative route for obtaining a chiroptical response of achiral structures is based on forming chiral arrangements between the incoming k-vector and the illuminated structure~\cite{c._w._bunn_chemical_1945,verbiest_optical_1996}. Intuitively, a 2D-symmetric split-ring resonator (SRR), side-lifted out of plane, has structural overlap with one loop of a helix~\cite{plum_metamaterials:_2009,singh_highly_2010,sersic_ubiquity_2012,kruk_spin-polarized_2014}. This so-called pseudo-chirality or extrinsic chirality is an inherently tunable effect in which the chiroptical response usually scales with the angle of incidence. Extrinsic chirality allows for control over the polarization properties of the transmitted light~\cite{singh_highly_2010} and optical activity of reflected waves~\cite{plum_specular_2016,plum_extrinsic_2016}. Spectrally-tunable enhanced extrinsic chirality was realized in flat metasurfaces, owing to the interaction of delocalized lattice surface modes with individual resonances, leading to strong narrowband circular dichroism (CD) spectra~\cite{leon_strong_2015} and spatially coherent circularly polarized fluorescence~\cite{cotrufo_spin-dependent_2016}. 
\\
The material composition of chiral nanostructures introduces an additional degree of freedom in tailoring the optical properties~\cite{yeom_chiral_2013,hentschel_babinet_2013,ogier_macroscopic_2014}. However, only recently the effect of heteromaterial selection was isolated~\cite{banzer_chiral_2016} by  showing differential absorption of a geometrically achiral nanostructure - a planar nanotrimer of three equally shaped nanodisks or spheres composed of different materials and assembled using a custom pick-and-place approach~\cite{bartenwerfer_towards_2011,mick_afm-based_2014,banzer_chiral_2016}. The chiroptical response of heterogeneous achiral nanostructures accentuates the capabilities of heterogeneous nanoscopic systems in chiral light-matter interactions. Particularly, a previous report~\cite{banzer_chiral_2016} suggests that heterogeneous morphology may lead to a chiroptical response in nanostructures possessing even simpler geometries.
\\
Here we theoretically and numerically~\cite{noauthor_fdtd_nodate} investigate the heterogeneity induced extrinsic chirality in a symmetric gold-silicon sphere heterodimer on a substrate. Subwavelength Si nanoparticles exhibit rich resonance spectra, consisting of the fundamental magnetic dipole resonance~\cite{craig_f._bohren_donald_r._huffman_absorption_1983,garcia-etxarri_strong_2011,evlyukhin_demonstration_2012} followed by the spectrally overlapping electric dipole and magnetic quadrupole resonances~\cite{wozniak_selective_2015}. The response of Au nanoparticles is dominated by the electric dipole resonance of plasmonic nature \cite{garcia_surface_2011}. In the case of the heterodimer, strong resonance hybridization~\cite{prodan_hybridization_2003,nordlander_plasmon_2004,muhlig_multipole_2011,zywietz_electromagnetic_2015,wang_janus_2015} is obtained if the incident polarization is aligned with the dimer axis (TE). For the case of the excitation field perpendicular to the dimer axis (TM), the scattering spectrum is  dominated by the Si nanoparticle response~\cite{wang_janus_2015}.
\\
Extrinsic chiral response is obtained for oblique incidence normal to the heterodimer axis. For symmetry reasons, no chiral response is expected for a homogeneous dimer~\cite{albella_low-loss_2013,wang_janus_2015} under oblique incidence, for a heterogeneous dimer under normal incidence, or in absence of a substrate. Hence, it is the coalescence of extrinsic chirality of the dimer-on-substrate and the heteromaterial morphology of the structure that imparts the chiroptical response.
\\
To quantify CD and circular birefringence (CB)~\cite{kuwata-gonokami_giant_2005,arteaga_relation_2016,hopkins_circular_2016} we adapt the M{\"u}ller matrix formalism~\cite{arteaga_o._mueller_2010} and k-space polarimetry~\cite{arteaga_complete_2014,osorio_k-space_2015}. We extend this formalism to account for oblique incidence and extrinsically chiral structures. Additionally, motivated by the demonstration of spin-polarized emission of fluorescent molecules and quantum dots coupled to extrinsically chiral nanostructures~\cite{kruk_spin-polarized_2014,cotrufo_spin-dependent_2016,yan_twisting_2017}, we study the interaction of the heterodimer~\cite{wang_janus_2015} with linearly polarized light (LP). We find that a spin-split behavior in scattering~\cite{bliokh_spin--orbital_2011} of LP is present in free space. Hence, spin-polarized scattering is a consequence of the heterogeneity of the structure alone and not of its extrinsic chirality.
\\
To further support these findings, we apply a coupled dipole model (CDM) for heterodimers~\cite{albella_low-loss_2013,albella_switchable_2015,shibanuma_unidirectional_2016} to quantify the dipole moments excited in each particle~\cite{banzer_chiral_2016}. The CDM reveals that heterogeneity results in excitation of transversely spinning~\cite{rodriguez-herrera_optical_2010,banzer_photonic_2013,oconnor_spinorbit_2014,aiello_transverse_2015,neugebauer_measuring_2015} electric and magnetic dipoles, along with a dipolar mode that resembles the emission of an SRR~\cite{rockstuhl_reinterpretation_2006,zhou_magnetic_2007,banzer_experimental_2010,muhlig_multipole_2011}. Both transversely spinning dipole and SRR modes emit spin-polarized light and the interplay of their relative strength and helicity leads to wavelength dependent spin-polarized scattering. The CDM combined with the substrate induced break of symmetry indicates that spin-polarized scattering is the origin of the extrinsic chirality.\\ \vspace{-1.0cm}

\section*{Results}
\vspace{-0.5cm}
\subsection*{Scattering cross-sections}
\begin{figure}[!htb]
\centering 
\includegraphics[width=0.48\textwidth]{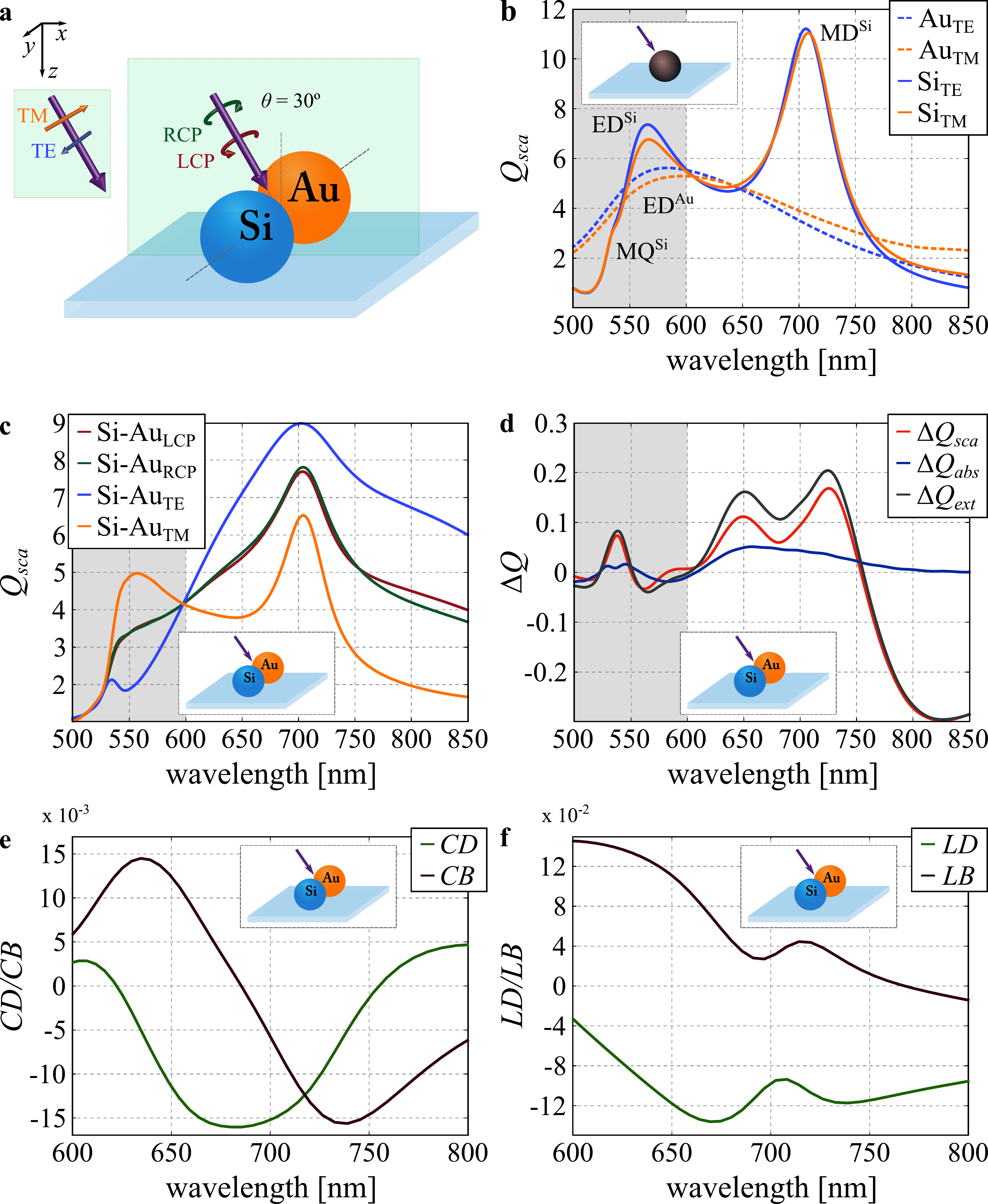} 
  \caption{$(\mathbf{a})$ Excitation scheme for the heterodimer on a glass substrate under oblique incidence ($\theta = 30^{\circ}$). $(\mathbf{b})$ Scattering efficiency spectra  for individual Au and Si nanoparticles ($\theta = 30^{\circ}$). Electric and magnetic dipolar (ED$\backslash$MD) and magnetic quadrupolar (MQ) resonances are indicated for Au and Si nanoparticles. $(\mathbf{c})$ Scattering efficiency spectra for the heterodimer in $(\mathbf{a})$. $(\mathbf{d})$ Differential scattering, absorption and extinction efficiency spectra for the heterodimer in $(\mathbf{a})$. $(\mathbf{e})$ Corresponding curves showing circular dichroism (CD), circular birefringence (CB) and $(\mathbf{f})$ linear dichroism (LD) and linear birefringence (LB) for the heterodimer in $(\mathbf{a})$. The insets indicate the excitation scheme. The shaded area in $(\mathbf{b})$-$(\mathbf{d})$ shows the spectral region with a strong quadrupolar contribution.\\ \vspace{-1.0cm}}
  \label{fig:_fig1}
\end{figure}

\vspace{-0.5cm}
For the numerical~\cite{noauthor_fdtd_nodate} demonstration of heterogeneity induced extrinsic chirality we choose a gold-silicon heterodimer consisting of spherical particles with radii $r=90\,\mathrm{nm}$ and distance $D = 182\,\mathrm{nm}$ between their centers, distributed along the $y$-axis and positioned on a glass substrate of refractive index $n=1.5$ (Fig.~\ref{fig:_fig1}$\mathbf{a}$). Fig.~\ref{fig:_fig1}$\mathbf{b}$ shows the numerically retrieved scattering efficiencies ($Q_{sca}$)~\cite{craig_f._bohren_donald_r._huffman_absorption_1983} for individual Si and Au nanoparticles positioned on a glass substrate, excited with linearly polarized light (TE$\backslash$TM with respect to the $xz$-plane of incidence) under oblique incidence at $\theta = 30^{\circ}$. Scattering efficiency is a dimensionless parameter measuring the total scattered power, normalized with respect to the incident intensity and the geometric cross section of the scatterer. For the individual Au and Si nanospheres notable differences in $Q_{sca}$ are observed between both excitation schemes due to the presence of a substrate~\cite{lukosz_light_1977,lukosz_light_1977-1,lukosz_light_1979,gay-balmaz_electromagnetic_2001,markovich_magnetic_2014,miroshnichenko_substrate-induced_2015} , which breaks the symmetry  between TE$\backslash$TM excitations under oblique incidence. For the case of the heterodimer, the heterogeneity breaks the mirror symmetry of the nanostructure, i.e. $xz$ is not a plane of mirror symmetry of the heterodimer. Combination of the heteromaterial-induced break of symmetry together with the geometrical break of symmetry by oblique incidence in the presence of a substrate results in different scattering, absorption and extinction efficiencies for incident right- and left-hand circular polarization (RCP$\backslash$LCP) (Fig.~\ref{fig:_fig1}$\mathbf{c}$). Fig.~\ref{fig:_fig1}$\mathbf{d}$ shows the resulting non-zero differential values $\Delta Q_{sca}$, $\Delta Q_{abs}$ and $\Delta Q_{ext}$ for the scheme indicated in Fig.~\ref{fig:_fig1}$\mathbf{a}$. Here,  $\Delta Q_{\alpha}=[Q_\alpha]_{RCP}-[Q_\alpha]_{LCP}$ are the scattering, absorption and extinction efficiencies, respectively, whereby the subscript of the square brackets indicates the incident polarization.\\ \vspace{-0.5cm}

\subsection*{M{\"u}ller matrix analysis}
\begin{figure}[!htb]
\centering
\includegraphics[width=0.48\textwidth]{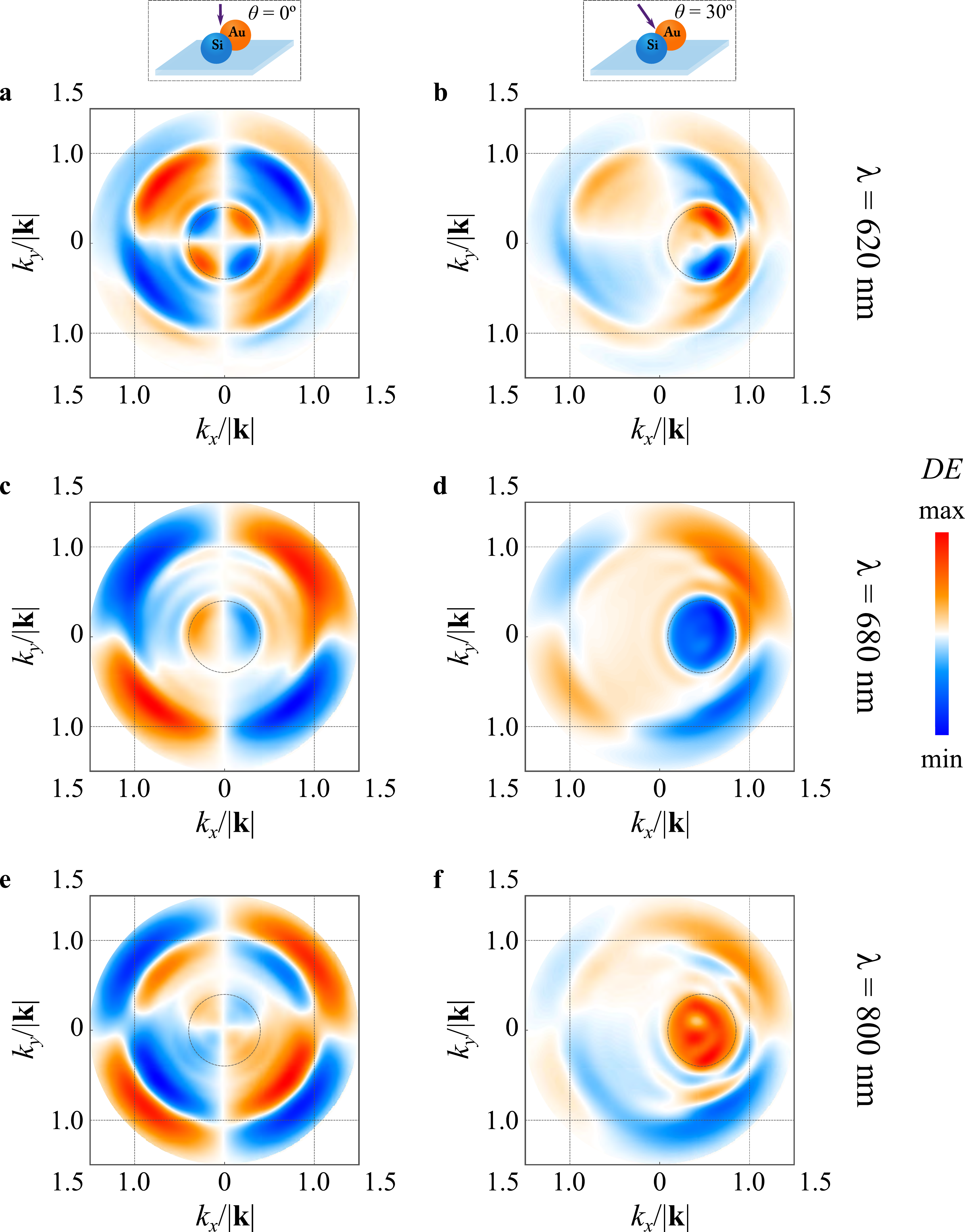} 
  \caption{Far-field distribution of differential extinction (DE) (Eq.~\ref{eq:_de}) for circularly polarized excitation at selected wavelengths. $(\mathbf{a})$,$(\mathbf{c})$,$(\mathbf{e})$ DE under normal incidence ($\theta = 0^{\circ}$) for the heterodimer on a glass substrate at $\lambda = 620, 680, 800\,\mathrm{nm}$, respectively. The integration area $\mathrm{NA} \leq 0.4$ is indicated. $(\mathbf{b})$,$(\mathbf{d})$,$(\mathbf{f})$  DE under oblique incidence at $\theta = 30^{\circ}$ and for the same wavelengths. The integration area $\mathrm{NA}  \leq 0.4$ for M{\"u}ller matrix analysis is squeezed and shifted to account for the refracted central wavevector. Each plot is normalized to its maximal value.\\ \vspace{-0.5cm}}
  \label{fig:_fig2}
\end{figure}

\vspace{-0.5cm}
For a detailed investigation of the polarimetric properties of the heterodimer-on-substrate under oblique incidence, we perform a M{\"u}ller matrix analysis in transmission, capable of separating the contributions of CD and CB from linear dichroism (LD) and linear birefringence (LB). To this end, the nanostructure is obliquely illuminated ($\theta = 30^{\circ}$) by a weakly focused Gaussian beam with effective numerical aperture $\mathrm{NA} =0.4$ and various incident polarization states $\boldsymbol{S}^{in}$, where $\boldsymbol{S}$ denotes the Stokes vector. Transmission into the substrate is collected in the far-field and recorded polarization resolved, from which the Stokes parameters are then derived and integrated for $\mathrm{NA} \leq0.4$. This method is formally similar to k-space polarimetry~\cite{arteaga_complete_2014,osorio_k-space_2015}, however the integration of Stokes parameters is done with respect to the refracted central wavevector for obliquely incident light and the adapted $\mathrm{NA} $ in transmission. The refraction and scaling of the $\mathrm{NA} $ results in the asymmetric integration area in the far-field plane shown in the right column of Fig.~\ref{fig:_fig2}. The M{\"u}ller matrix $\boldsymbol{M}^{num}$ of the heterodimer-on-substrate under oblique incidence ($\theta = 30^{\circ}$) is obtained from the angularly integrated output Stokes vectors $\boldsymbol{S}^{out}$ by inverting the identity $\boldsymbol{S}^{out}=\boldsymbol{M}^{num}\boldsymbol{S}^{in}$ for a set of incident polarizations $\boldsymbol{S}^{in}$~\cite{arteaga_o._mueller_2010}. We utilize six input polarizations: RCP, LCP, TE, TM, diagonal and antidiagonal polarizations with respect to the TE$\backslash$TM basis. Next, we normalize the M{\"u}ller matrix $\boldsymbol{M}^{num}$ by its upper-left element $\boldsymbol{M}=\boldsymbol{M}^{num}/m_{00}^{num}$ and apply Cloude\textquoteright s sum decomposition~\cite{cloude_conditions_1990} to obtain a non-depolarizing estimate $\boldsymbol{M}^{nd}$ of $\boldsymbol{M}$. The depolarization arises due to a finite integration region $\mathrm{NA} \leq0.4$ shown in Fig.~\ref{fig:_fig2}. Finally, we calculate $\mathrm{CD} =0.5  [m_{03}^{nd}+m_{30}^{nd}]$, $\mathrm{CB} =0.5  [m_{12}^{nd}-m_{21}^{nd}]$, $\mathrm{LD} =-0.5  [m_{01}^{nd}+m_{10}^{nd}]$ and $\mathrm{LB} =0.5  [m_{32}^{nd}-m_{23}^{nd}]$, where $m_{ij}^{nd}$ is element of $\boldsymbol{M}^{nd}$. For clarity, we restrict ourselves to the wavelength range $600 \,\mathrm{nm} \leq \lambda \leq 800 \,\mathrm{nm}$ around the magnetic dipole resonance of the Si nanoparticle and an angle of incidence of $\theta = 30^{\circ}$. Fig.~\ref{fig:_fig1}$\mathbf{e}$ shows that the nanodimer exhibits CD and CB, almost perfectly matching the Born-Kuhn dispersion~\cite{yin_interpreting_2013}. In addition we present the significance of LD and LB in Fig.~\ref{fig:_fig1}$\mathbf{f}$, originating from the anisotropy of the structure. It is worth noting here that LD and LB are the only non-zero components for normally incident illumination (not shown here).\\ \vspace{-1.0cm}

\subsection*{Angularly resolved differential extinction}
\vspace{-0.5cm}
The symmetry properties of the heterodimer-on-substrate can be illustrated by plotting differential extinction (DE) of RCP$\backslash$LCP in an angularly resolved fashion, defined as 
\begin{equation}
\mathrm{DE} (k_{x},k_{y})=\frac{1}{2} [S_{0}+S_{3}]_{RCP}-\frac{1}{2}[S_{0}-S_{3}]_{LCP} \text{,} 
\label{eq:_de}
\end{equation}
where the Stokes parameters $S_{i} (k_{x},k_{y})$ are measured in the far-field as a function of k-space coordinates and the subscript indicates the incident polarization. The first and second term of the right-hand side of Eq.~\ref{eq:_de} are the projections of the output Stokes vector on RCP$\backslash$LCP~\cite{michael_bass_casimer_decusatis_jay_m._enoch_vasudevan_lakshminarayanan_guifang_li_carolyn_macdonald_virendra_n._mahajan_eric_van_stryland_handbook_2009}, respectively. Fig.~\ref{fig:_fig2} shows the angularly resolved far-field DE data for normal (left column) and oblique ($\theta = 30^{\circ}$, right column) incidence at three different wavelengths, normalized to its maximum value, respectively. It should be noted that only within $\mathrm{NA} \leq0.4$ Eq.~\ref{eq:_de} represents differential extinction, while outside of this angular range it represents the differential scattering. The left column of Fig.~\ref{fig:_fig2} clearly shows that the symmetry of the system with respect to  $\pm k_{y}$ (top to bottom) is broken for normal incidence due to the heterogeneity. However, the symmetry with respect to $\pm k_{x}$ is preserved, owing to the $yz$-plane of mirror symmetry, resulting in  $\mathrm{DE} (k_{x},k_{y})=-\mathrm{DE} (-k_{x},k_{y})$ at $\theta = 0^{\circ}$. Oblique incidence breaks the latter symmetry, resulting in non-zero DE within the marked area of $\mathrm{NA} \leq0.4$. Fig.~\ref{fig:_fig2}$\mathbf{d}$ and Fig.~\ref{fig:_fig2}$\mathbf{f}$ show negative and positive $\mathrm{DE} (k_{x},k_{y})$ within $\mathrm{NA} \leq0.4$ for $\lambda = 680 \,\mathrm{nm}$ and $\lambda = 800 \,\mathrm{nm}$, corresponding to the minimal and maximal value of the CD curve in Fig.~\ref{fig:_fig1}$\mathbf{e}$, respectively. Nevertheless, the integrated value of $\mathrm{DE} (k_{x},k_{y})$ may be zero under oblique incidence, as it is the case for $\lambda = 620 \,\mathrm{nm}$ (Fig.~\ref{fig:_fig2}$\mathbf{b}$). Please note that also the CD is zero at this wavelength as shown in Fig.~\ref{fig:_fig1}$\mathbf{e}$.\\ \vspace{-1.0cm}

\subsection*{Spin-polarized scattering}
\vspace{-0.5cm}
The odd symmetry of $\mathrm{DE} (k_{x},k_{y})$ with respect to $k_{x}$ in the left column of Fig.~\ref{fig:_fig2} suggests that the heterodimer exhibits spin-polarized scattering for normally incident LP, an effect that is also responsible for the spin-polarized emission of extrinsically chiral structures in the presence of fluorescent molecules or quantum dots~\cite{kruk_spin-polarized_2014,cotrufo_spin-dependent_2016,yan_twisting_2017}. To quantify the spin-polarized scattering of normally incident LP we define the dissymmetry factors $\Delta_{TE}$ and $\Delta_{TM}$ as $\Delta_{\alpha}=[\widetilde{S}_{3}(k_{x}>0)-\widetilde{S}_{3}(k_{x}<0)]_{\alpha}$, where $\widetilde{S}_{3}(k_{x} \lessgtr 0)$ is the integrated forward $(z \geq 0)$ scattered  far-field $S_{3}$ parameter in units of scattering efficiency. Fig.~\ref{fig:_fig3}$\mathbf{a}$ shows that the heterodimer excited with  linearly TE$\backslash$TM polarized light at normal incidence scatters in a pronounced spin-split manner. Fig.~\ref{fig:_fig3}$\mathbf{c}$ and \ref{fig:_fig3}$\mathbf{d}$ exemplarily show the distributions of scattered $S_{3}(k_{x},k_{y})$ for TE excitation at $\lambda = 640 \,\mathrm{nm}$ and $\lambda = 720\,\mathrm{nm}$, normalized to the maximal value of scattered intensity $S_{0}$. Fig.~\ref{fig:_fig3}$\mathbf{b}$. shows the decomposition of the scattering efficiency into different polarization components, $Q_{TE(tot)}$, $Q_{TE \rightarrow  LP}$ and $Q_{TE \rightarrow CP}$, for normally incident TE polarization, i.e., integrated $S_{0}$, $\sqrt{S_{1}^{2}+S_{2}^{2}}$ and $\left|S_{3}\right|$ in units of scattering efficiency. Interestingly, $\Delta_{TE}$ and $\Delta_{TM}$ are of the same order of magnitude and of opposite sign, which has two important consequences. First, the heterodimer is not optimized to achieve maximal chiral response, with its parameters studied here, hence having a potential for improvement. Second, since RCP$\backslash$LCP are the superpositions of TE$\backslash$TM with $\pm \frac{\pi}{2}$ phase difference and $\Delta_{TE}\approx - \Delta_{TM}$, the scattered light for incident RCP$\backslash$LCP will show a strong contribution of linear polarization. Indeed, in Fig.~\ref{fig:_fig3}$\mathbf{e}$ we plot the decomposition of the scattering efficiency into different polarization components for normally incident RCP light, $Q_{RCP(tot)}$, $Q_{RCP \rightarrow  LP}$ and $Q_{RCP \rightarrow CP}$, revealing that for most wavelengths in the investigated spectral range the scattered LP prevails. This is a consequence of the strong  anisotropy and the resulting LD and LB spectra shown in Fig.~\ref{fig:_fig1}$\mathbf{f}$. Notably, the $\widetilde{S}_{3}$ for incident LP in free space is zero if the incident polarization is aligned with the symmetry axes of the heterodimer, i.e. TE$\backslash$TM. This means that the overall chirality flux (CF)~\cite{poulikakos_optical_2016} generated by the structure in forward direction is zero. In the presence of a substrate this phenomenon is only to be observed for normal incidence. For obliquely incident TE$\backslash$TM polarized light $\widetilde{S}_{3}$ is not zero. This also explains the different extinction spectra for the incident RCP$\backslash$LCP components presented in Fig.~\ref{fig:_fig1}$\mathbf{c}$ and~\ref{fig:_fig1}$\mathbf{d}$. Fig.~\ref{fig:_fig3}$\mathbf{f}$ shows the generated CF, i.e. $\widetilde{S}_{3}$ in units of scattering efficiency, for $\theta = 30^{\circ}$. Owing to the difference in the generated CF for TE$\backslash$TM incident light, it is clear that only the complete M{\"u}ller matrix analysis presented earlier is capable of separating the contributions of CD and CB from LD and LB.\\ \vspace{-1.0cm}

\subsection*{Dipolar modes of the heterodimer}
\begin{figure}[!htb]
\centering
\includegraphics[width=0.48\textwidth]{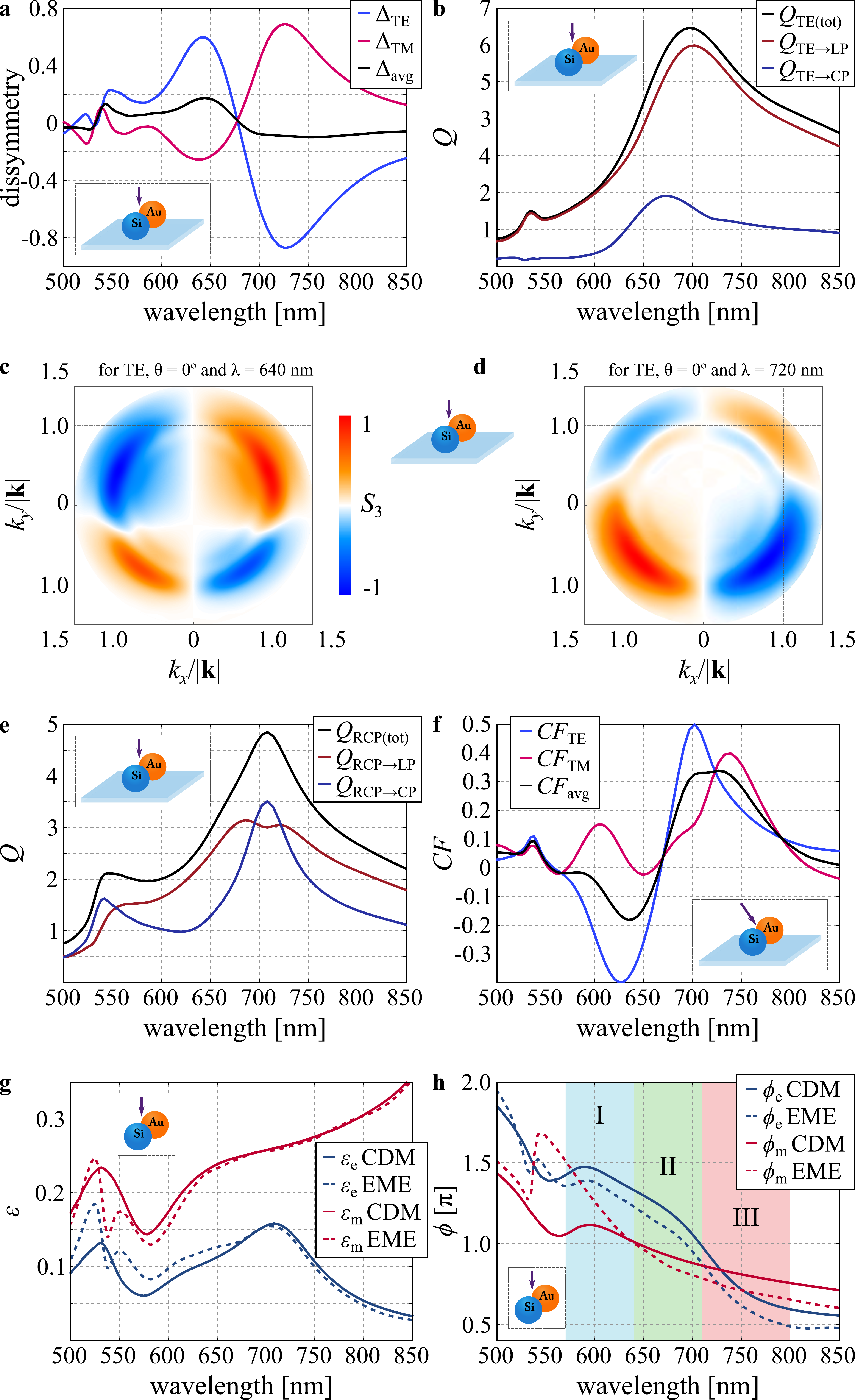} 
  \caption{$(\mathbf{a})$ Dissymmetry factors of scattered circularly polarized component under normally incident TE$\backslash$TM polarized light. $(\mathbf{b})$ Decomposition of scattering efficiency for TE polarized light normally incident on the heterodimer into different polarization components. $(\mathbf{c})$,$(\mathbf{d})$ Far-field spin-split scattering for normally incident TE polarized light at $\lambda= 640,720\,\mathrm{nm}$, respectively. $(\mathbf{e})$ Decomposition of scattering efficiency for normally incident RCP into different polarization components. $(\mathbf{f})$ Chirality flux (CF) in forward direction for obliquely incident TE$\backslash$TM polarized light at $\theta = 30^{\circ}$. $(\mathbf{g})$ Eccentricity of the total transversely spinning dipole, $\upvarepsilon_{e}$, and split-ring-resonator, $\upvarepsilon_{m}$, modes calculated using coupled dipole model (CDM) and exact multipole expansion (EME) for the heterodimer in free space and their corresponding phases $\phi_{e}$ and $\phi_{m}$. $(\mathbf{h})$ The regions with positive (blue), negative (red) and opposite (green) helicities are highlighted. The insets indicate the excitation scheme.\\\vspace{-1.0cm}}
  \label{fig:_fig3}
\end{figure}

\vspace{-0.5cm}
There is a striking difference between the observed spin-polarized scattering in Fig. 3c and 3d of the heterodimer and the one from extrinsically chiral nanostructures~\cite{kruk_spin-polarized_2014,cotrufo_spin-dependent_2016,yan_twisting_2017}. While the heterodimer is not extrinsically chiral in free space, it still scatters light in spin-split manner. Hence, the exclusive reason for the spin-polarized scattering is the heterogeneity. The effect can be understood from symmetry considerations~\cite{banzer_chiral_2016} based on a coupled dipole model~\cite{albella_low-loss_2013,albella_switchable_2015}. Consider TE polarized light incident along the $z$-direction, exciting the dipolar response in each particle of the heterodimer in free space (no substrate). For individual particles, neglecting the mutually induced dipoles, the electric dipole moments $p_{y}^{Au}$, $p_{y}^{Si}$ and the magnetic dipole moment $m_{x}^{Si}$ are exited. The magnetic dipole moment $m_{x}^{Si}$ in the Si nanoparticle excites a phase-delayed longitudinal dipole $p_{z}^{Au}$ in the Au nanoparticle, which in turn induces a phase-delayed longitudinal dipole moment $p_{z}^{Si}$ in Si~\cite{albooyeh_purely_2016}. Hence,incident TE light excites transversely spinning dipoles in both, the Au and the Si nanoparticle. The spinning plane of such a dipole splits the space in two halves, and a transversely spinning dipole emits light of opposite helicity in each of those half spaces~\cite{rodriguez-herrera_optical_2010,oconnor_spinorbit_2014}. This effect is also known as the giant spin Hall effect of light~\cite{rodriguez-herrera_optical_2010}. Excitation of transversely spinning dipoles in symmetric nanoparticles requires complex beams~\cite{rodriguez-herrera_optical_2010,banzer_photonic_2013,neugebauer_polarization_2014,neugebauer_measuring_2015} or a substrate and obliquely incident CP~\cite{oconnor_spinorbit_2014}. In the case presented here, the transversely spinning dipoles are excited for a plane-wave excitation linearly polarized along the axis of symmetry of the heterodimer~\cite{albooyeh_purely_2016}. Moreover, a combination of $m_{x}$ and the dephased $p_{z}$ in each particle also scatters light in a spin-polarized fashion. This combination $(m_{x},p_{z})$ with a phase delay of $\phi = \pm \frac{\pi}{2}$ is equivalent to the fundamental mode of an SRR~\cite{rockstuhl_reinterpretation_2006,zhou_magnetic_2007,banzer_experimental_2010,muhlig_multipole_2011} located in the $yz$-plane (cf. Fig.~\ref{fig:_fig1}$\mathbf{a}$) with arms pointing along the $y$-axis. This unbalanced interplay between transversely spinning dipoles and SRR modes in the Au and Si nanoparticles of different strength and helicity  leads to wavelength dependent spin-split scattering (Fig.~\ref{fig:_fig3}$\mathbf{c}$ and~\ref{fig:_fig3}$\mathbf{d}$). Same arguments apply for incident TM polarized light with the roles of electric and magnetic field interchanged.\\ \vspace{-1.0cm}

\subsection*{Coupled dipole model}
\vspace{-0.5cm}
To quantify the spin-polarized scattering in free space we use a coupled dipole model. We calculate the excited electric and magnetic dipolar moments in the individual particles under linearly polarized TE illumination and plot the relative phase and eccentricity of the overall transversely spinning dipole and SRR modes defined as:
\begin{subequations}
\begin{align}
\upvarepsilon_{e}&=\frac{2}{\pi} atan\left\{  \left|   \frac{p_{z}^{Si}+p_{z}^{Au}}{p_{y}^{Si}+p_{y}^{Au}}  \right| \right\}\text{,}
\label{eq:_epe}\\
\upvarepsilon_{m}&=\frac{2}{\pi} atan\left\{  \left|   c\frac{p_{z}^{Si}+p_{z}^{Au}}{m_{x}^{Si}+m_{x}^{Au}}  \right| \right\}\text{,}
\label{eq:_epm}\\
\phi_{e}&= Arg \left\{     \frac{p_{z}^{Si}+p_{z}^{Au}}{p_{y}^{Si}+p_{y}^{Au}}   \right\}\text{,}
\label{eq:_fie}\\
\phi_{m}&=Arg \left\{     -\frac{p_{z}^{Si}+p_{z}^{Au}}{m_{x}^{Si}+m_{x}^{Au}}   \right\} \text{.}
\label{eq:_fim}
\end{align}
\end{subequations} 
Here $\upvarepsilon_{m},\upvarepsilon_{e}$, are the eccentricities of the total transversely spinning dipole and the SRR modes, $\phi_{m},\phi_{e}$ are thier phases and $c$ is the speed of light in vacuum. The induced magnetic dipole moment in the gold nanoparticle $m_{x}^{Au}$ must be included in the model, since $\left|\frac{m_{x}^{Au}/c}{p_{y}^{Au}}\right| \sim 10^{-1}$ in the investigated spectral range. Fig.~\ref{fig:_fig3}$\mathbf{g}$ and \ref{fig:_fig3}$\mathbf{h}$ show the results of Eq.~\ref{eq:_epe} -~\ref{eq:_fim} obtained by the  CDM~\cite{albella_low-loss_2013,albella_switchable_2015} and from the exact multipole expansion (EME)~\cite{alaee_electromagnetic_2018} of the displacement currents in each nanoparticle. The discrepancies between CDM and EME originate from the quadrupolar modes excited in the Si nanoparticle at shorter wavelengths. This spectral range is shaded in  Fig.~\ref{fig:_fig1}$\mathbf{b}$ -~\ref{fig:_fig1}$\mathbf{d}$. Both transversely spinning dipole and SRR modes are excited and show non-zero eccentricity in the whole spectral range (Fig.~\ref{fig:_fig3}$\mathbf{g}$). For the relative phases $\phi_{m},\phi_{e}$, Fig.~\ref{fig:_fig3}$\mathbf{h}$ shows three distinct regions, colorized blue, green and red, respectively, and summarized in Table~\ref{tab:_tab1}. In each of the regions I and III the total transversely spinning dipole and SRR modes have the same positive or negative helicity, $\sigma_{e,m} \equiv sign\left\{sin⁡(-\phi_{e,m})\right\}$, respectively. Their spin-split scattering increases the absolute value of the dissymmetry factor $\Delta_{TE}$. In region II, the transversely spinning dipole and SRR modes have opposite helicity and their interference decreases the absolute value of $\Delta_{TE}$. These features are consistent with the dissymmetry factor $\Delta_{TE}$ for the heterodimer-on-substrate in Fig.~\ref{fig:_fig3}$\mathbf{a}$, showing positive (negative) values in region I (III). The corresponding results for the incident TM polarized light may be obtained via duality transformation~\cite{novotny_principles_2012} and the effect of a substrate can be accounted for in CDM by following the derivations in references~\cite{albella_low-loss_2013,albella_switchable_2015} with Green\textquoteright s tensor for stratified media~\cite{paulus_greens_2001}.\\
A dipolar scatterer positioned at a sub-wavelength distance above an interface (substrate) emits mostly into the medium with higher optical density~\cite{lukosz_light_1977,lukosz_light_1977-1,lukosz_light_1979,novotny_principles_2012} (Fig.~\ref{fig:_fig1}$\mathbf{a}$, $z>0$). Consequently, for oblique incidence on a heterodimer positioned on a substrate, CP of a preferred helicity will couple more efficiently in forward direction. This explains the non-zero generation of CF (Fig.~\ref{fig:_fig3}$\mathbf{f}$) and, eventually, the extrinsic chirality of the heterodimer-on-substrate arrangement. Hence, the CDM, combined with the substrate induced break of symmetry, reveals the relation between spin-polarized scattering and the origin of the presented extrinsic chiroptical response.
\begin{table}[ht]
\centering

\begin{tabular}{||c|c|c||} 
\hline
Region& Mode Helicity&Wavelelgnths\\

\hline\hline
I&$\sigma_{e},\sigma_{m}>0$& $570 \leq \lambda < 640$ \\

\hline
II&$\sigma_{e}>0,\sigma_{m}<0$ & $640 \leq \lambda < 710$ \\

\hline
III& $\sigma_{e},\sigma_{m}<0$ & $710 \leq \lambda \leq 800$ \\
\hline

	\end{tabular}
  \caption{Summary of the regions highlighted in Fig.~\ref{fig:_fig3}$\mathbf{h}$. In region I and III the transversely spinning dipole and SRR modes have same helicity. In region II they have opposite helicity.\\ \vspace{-1.5cm}}
  \label{tab:_tab1}
\end{table}

\section*{Conclusion}
\vspace{-0.5cm}
In conclusion, we have investigated a geometrically symmetric heterogeneous sphere dimer on a substrate. This structure exhibits extrinsic chirality only in the presence of a substrate, owing to the heterogeneous morphology. We characterized the transmission properties of the structure by extended k-space polarimetry, to account for oblique incidence. The M{\"u}ller matrix analysis reveals the chiroptical response of the heterodimer-on-substrate with respect to circular dichroism and circular birefringence and much stronger linear dichroism and linear birefringence. This strong inherent anisotropy of the structure leads to a pronounced conversion of the incident circularly polarized light to scattered linearly polarized light. A coupled dipole model analysis of spin-polarized scattering of incident linearly polarized light reveals the transversely spinning dipole and SRR dipolar modes excited in the heterodimer in free space. The unbalanced spectral interplay of these modes together with the substrate induced break of symmetry elucidate the physical origins of extrinsic chirality of the heterodimer-on-substrate. Our results provide new insights into the interaction of light with heterogeneous nanostructures and spin-polarized scattering, emphasizing the capabilities and potential of heterogeneous nanoparticle systems tailored at the nanoscale. In addition, our findings may constitute a novel route towards the realization of material-tailored polarization selective metasurfaces.\\ \vspace{-1.0cm}

\begin{acknowledgments}
\vspace{-0.5cm}
The authors gratefully acknowledge fruitful discussions with Israel De Leon, Rasoul Alaee, Oriol Arteaga and Gerd Leuchs.
\end{acknowledgments}
\bibliography{bibliography/Manusc1_Dimer_Ext_Chir}
\end{document}